\documentclass{aa}

\input epsf 

\begin{document}

\sloppy

\thesaurus{3 (11.02.2, 13.07.2)} 

\title{Reanalysis of the high energy cutoff of the 1997 Mkn 501 
TeV energy spectrum}

\titlerunning{Reanalysis of the Mkn 501 Energy Spectrum}
\authorrunning{F. Aharonian et al.}

\author{F.A. Aharonian\inst{1},
A.G.~Akhperjanian\inst{7},
J.A.~Barrio\inst{2,3},
K.~Bernl\"ohr\inst{1},
O.~Bolz\inst{1},
H.~B\"orst\inst{5},
H.~Bojahr\inst{6},
J.L.~Contreras\inst{3},
J.~Cortina\inst{2},
S.~Denninghoff\inst{2}
V.~Fonseca\inst{3},
J.C.~Gonzalez\inst{3},
N.~G\"otting\inst{4},
G.~Heinzelmann\inst{4},
G.~Hermann\inst{1},
A.~Heusler\inst{1},
W.~Hofmann\inst{1},
D.~Horns\inst{4},
A.~Ibarra\inst{3},
C.~Iserlohe\inst{6},
I.~Jung\inst{1},
R.~Kankanyan\inst{1,7},
M.~Kestel\inst{2},
J.~Kettler\inst{1},
A.~Kohnle\inst{1},
A.~Konopelko\inst{1,}$^\S$,
H.~Kornmeyer\inst{2},
D.~Kranich\inst{2},
H.~Krawczynski\inst{1},
H.~Lampeitl\inst{1},
E.~Lorenz\inst{2},
F.~Lucarelli\inst{3},
N.~Magnussen\inst{6},
O.~Mang\inst{5},
H.~Meyer\inst{6},
R.~Mirzoyan\inst{2},
A.~Moralejo\inst{3},
L.~Padilla\inst{3},
M.~Panter\inst{1},
R.~Plaga\inst{2},
A.~Plyasheshnikov\inst{1,}$^\S$,
J.~Prahl\inst{4},
G.~P\"uhlhofer\inst{1},
W.~Rhode\inst{6},
A.~R\"ohring\inst{4},
G.P.~Rowell\inst{1},
V.~Sahakian\inst{7},
M.~Samorski\inst{5},
M.~Schilling\inst{5},
F.~Schr\"oder\inst{6},
M.~Siems\inst{5},
W.~Stamm\inst{5},
M.~Tluczykont\inst{4},
H.J.~V\"olk\inst{1},
C.~Wiedner\inst{1},
W.~Wittek\inst{2}}

\institute{Max Planck Institut f\"ur Kernphysik,
Postfach 103980, D-69029 Heidelberg, Germany \and
Max Planck Institut f\"ur Physik, F\"ohringer Ring
6, D-80805 M\"unchen, Germany \and
Universidad Complutense, Facultad de Ciencias
F\'{i}sicas, Ciudad Universitaria, E-28040 Madrid, Spain 
\and
Universit\"at Hamburg, II. Institut f\"ur
Experimentalphysik, Luruper Chaussee 149,
D-22761 Hamburg, Germany \and
Universit\"at Kiel, Institut f\"ur Experimentelle und Angewandte Physik,
Leibnizstra{\ss}e 15-19, D-24118 Kiel, Germany\and
Universit\"at Wuppertal, Fachbereich Physik,
Gau{\ss}str.20, D-42097 Wuppertal, Germany \and
Yerevan Physics Institute, Alikhanian Br. 2, 375036 Yerevan, 
Armenia\\
\hspace*{-4.04mm} $^\S\,$ On leave from  
Altai State University, Dimitrov Street 66, 656099 Barnaul, Russia\\
}

\mail{Werner Hofmann, \\Tel.: (Germany) +6221 516 330,\\
email address: Werner.Hofmann@mpi-hd.mpg.de}

\offprints{Werner Hofmann}

\date{Received 18 August 2000; accepted }

\maketitle

\begin{abstract}

Data taken with the HEGRA system of imaging atmospheric 
Cherenkov telescopes during the 1997 flares of Markarian 501
(Mkn 501)
are reanalyzed using an algorithm providing improved 
energy resolution. A resolution of 10\% to 12\% is obtained
by accounting for the variation of the Cherenkov light 
yield with the height of the shower maximum in the atmosphere. The improved
energy resolution is particularly relevant for the study of the high-energy
cutoff in the spectrum, which might be caused by interactions
with the intergalactic infrared background radiation.
The reanalysis presented here confirms the results obtained in the previous
analysis, but hints a steeper slope of the spectrum in 
the region around 20~TeV.


\end{abstract}

\section{Introduction}

The attenuation of extragalactic gamma rays in the TeV energy range
due to interactions
with the cosmic infrared background radiation
(Nikishov 1962; Gould \& Schr\'eder 1967; Stecker et al. 1992) 
is under intense discussion
(see, e.g., Funk et al. 1998; Stecker \& de Jaeger 1998; 
Aharonian et al. 1999a; Coppi and Aharonian 1999;
Konopelko et al. 1999a; Stecker 1999; Krawczynski et al. 2000; Vassiliev 2000;
and references given therein),
in particular since the energy spectra determined for the AGN Mkn 501
(Aharonian et al. 1999a,b,c;
Samuelson et al. 1998; Djannati-Atai et al. 1999),
obtained during its 1997 flares, show indications of absorption
features, particularly evident in the high-energy HEGRA data
of Aharonian et al. (1999a),
in the following referred to as Paper I. 
\begin{figure*}[htb]
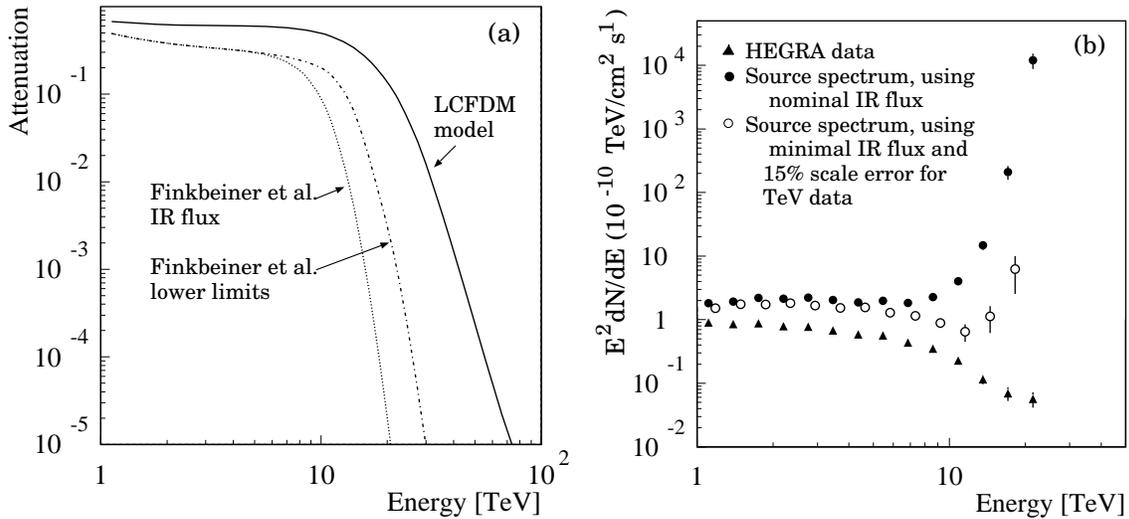

\begin{center}
\mbox{
\epsfxsize7.5cm
\epsffile{h2398f1a.eps}}
\mbox{
\epsfxsize7.2cm
\epsffile{h2398f1b.eps}}
\label{fig_abs}
\caption{(a) Attenuation of TeV gamma rays from Mkn 501 due
to interaction with the infrared background radiation, for
different assumptions concerning the density of the
infrared background. Dotted line: following recent data
from Finkbeiner et al. (2000), 
as shown by the solid line of Fig. 1 in Protheroe \& Meyer (2000).
Dashed-dotted line: using the minimal flux consistent with 
the $1 \sigma$ statistical and systematic error bars of 
Finkbeiner et al. (2000).
Full line: earlier LCFDM model of Primack et al. (1999).
All curves use are calculated for the redshift
$z \approx 0.03$ of Mkn 501, assuming a Hubble constant
of 65 km s$^{-1}$Mpc$^{-1}$.
(b) Spectral energy distribution
$E^2$d$N$/d$E$
of TeV gamma rays from Mkn 501 at the source, reconstructed
by dividing the measured distribution (HEGRA, Paper I, shown as
triangles) by the
attenuation. Filled circles: using the attenuation indicated by the
dotted curve in part (a) of this Figure. Open circles: using the 
attenuation indicated by the dashed-dotted curve, shifting
the HEGRA data down in energy by 15\% -- the maximal systematic
error on the energy scale -- and including the systematic error bars
of the HEGRA measurement.}
\end{center}
\end{figure*}
The level of the cosmic infrared background is related to
the evolution of galaxies and can serve to distinguish different
cosmological models
(see, e.g., Primack et al. 1999). 
However, due to the intense foreground radiation, a 
reliable direct measurement of the infrared background is a difficult
task. Gamma rays in the TeV energy range 
provide an independent and complementary approach
to study the infrared background radiation. Since the $\gamma \gamma 
\rightarrow e^+e^-$
interaction cross section peaks near threshold, the absorption of
gamma rays of energy $E$ probes the background level around a wavelength
$\lambda/\mu\mbox{m} \approx \kappa E/\mbox{TeV}$, where
$\kappa$ varies between 1 and 2 depending on the
spectral shape of the background radiation. 
There is almost a one-to-one correspondence between the energy-dependent
attenuation and the spectral density of the infrared background radiation.
Recent analyses by Finkbeiner et al. (2000) of the background 
level at long wavelengths, at 60~$\mu$m and 100~$\mu$m,
show a significantly higher flux than expected, which results
in strong absorption of Mkn 501 gamma rays 
above 10 TeV (Fig.~\ref{fig_abs}(a)). While earlier analyses
using a lower level of the infrared background 
(e.g., the models of Primack et al. (1999), also illustrated in Fig.~\ref{fig_abs}(a))
found good consistency
with the TeV data (see, e.g., Konopelko et al. (1999a)), 
these new results appear to be in contradiction with measurements
using the HEGRA Cherenkov telescopes,
which show a gamma-ray flux from Mkn 501 up to at least 16~TeV
(Paper I).
In order to explain the observed flux in the presence of
strong high-energy attenuation, the spectral distribution
of gamma-rays at the source would have to exhibit a rise at high
energies (Fig.~\ref{fig_abs}(b)), 
inconsistent both with acceleration models and with energetics
of the source
(see, e.g., Paper I and 
Coppi and Aharonian 1999;
Guy et al. 2000;
Protheroe \& Meyer 2000;
Finkbeiner et al. 2000; and references
given therein). 
Exotic solutions of this ``gamma-ray crisis'' have
been proposed, such as violations of Lorentz invariance 
(Kifune 1999) or photon
condensates (Harwit et al. 1999); 
the latter explanation has been questioned on rather general
grounds (Levinson 2000) and seems already to be ruled out by a recent
HEGRA analysis (Aharonian et al. 2000). 

The TeV gamma-ray data and the infrared background data
can be made marginally consistent by fully exploiting the 
systematic errors of both measurements, i.e.,
taking the minimal infrared flux values and gamma ray flux values
and reducing the HEGRA energy
scale by 15\%. 
In this case, a marginally acceptable source spectrum is achieved, see
Fig.~\ref{fig_abs}(b), in particular given that the highest HEGRA data
point is, at the $2\sigma$ level, consistent with spill-over
from the lower energy bins. The bulk of the change in the calculated
source spectrum -- a factor 100 for the highest data point --
can be attributed to the uncertainty in the infrared
data; the effect of the HEGRA uncertainties is smaller -- about a factor
10 for the highest point.

The interpretation of the experimental results is currently
limited by the systematic errors both of the measurement of
the infrared background, and of the TeV spectra. One 
potentially very
important factor is the energy
resolution in the detection of the TeV gamma rays. 
The HEGRA system of Cherenkov telescopes, which
provides the highest-energy data, has an energy resolution around
20\% (see Aharonian et al. 1999b, Konopelko et al. 1999b,
Paper I). While this resolution is ample for the reconstruction of the usual 
power-law energy spectra, it may smear out the extremely
sharp cutoff features expected for absorption due to infrared 
background, in particular for high long-wavelength infrared background levels
(Fig.~\ref{fig_abs}(a)). The highest-energy gamma rays detected
could simply represent spill-over from lower energies
due to the finite energy resolution. Indeed,
while Aharonian et al. (Paper I) 
measure photons with reconstructed energies up
to 21.5 TeV and beyond, they state that a 
sharp cutoff at 16 TeV cannot be excluded by the data.
It is obviously of considerable interest to improve the 
energy reconstruction of the system of telescopes, in order to more
precisely map a possible cutoff. A scheme to improve the energy resolution
of systems of Cherenkov telescopes by roughly a factor 2 
has been discussed by Hofmann et al. (2000), making use of
the abundant information provided by such systems for each 
individual gamma-ray air shower. In this note, we present
a reanalysis of the HEGRA Mkn 501 spectral data with such improved
analysis algorithms.

\section{Determination of gamma ray spectra}

One of the key limiting factors in the energy determination of
air showers with imaging Cherenkov telescopes is the fluctuation
in the height of the shower maximum in the atmosphere. 
Deeper showers have a larger
light yield both because of the reduced distance to the telescopes
and because of the reduction of the threshold for Cherenkov light emission
with
decreasing height, which allows more particles to contribute to
the emission of Cherenkov light. As demonstrated by Hofmann et al. 
(2000)
and by Aharonian et al. (2000),
 the multiple views of a shower
provided by the HEGRA stereoscopic system of Cherenkov telescopes 
allow to determine the height of the shower maximum with an error
of 500~m to 600~m. Using look-up tables which depend both on the distance to
the shower core and the shower height to relate the Cherenkov light yield to the 
gamma-ray energy, the energy resolution of the HEGRA system can be
improved significantly (Hofmann et al. 2000). A further improvement
can be reached, if the known location of the gamma-ray source is used
as input in the reconstruction of the shower geometry, providing a 
greatly improved localization of the shower core, and hence a better
measurement of the distance of a given telescope to the shower core.
Combining all these techniques, an energy resolution around 10\% to 12\% is
obtained (Hofmann et al. 2000), 
implying - for the identical set of events -
a factor 2 improvement over more conventional
techniques such as used in Aharonian et al. (1999b) and in Paper~I.

\begin{figure}[htb]
\begin{center}
\mbox{
\epsfxsize7.5cm
\epsffile{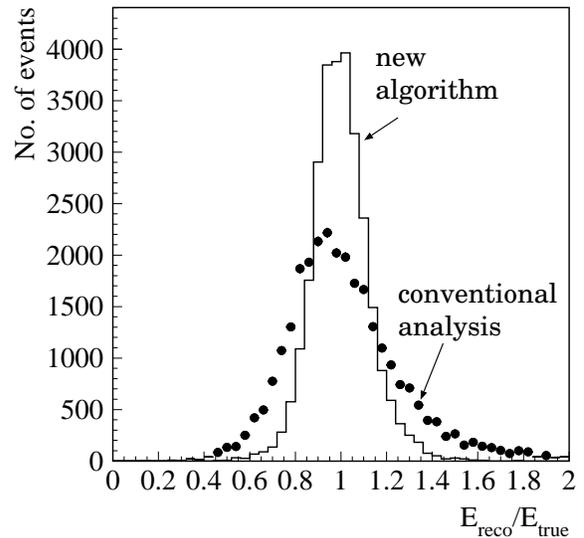}}
\label{fig_eres}
\caption{Ratio of the reconstructed and true gamma-ray energies, 
as derived from Monte-Carlo simulations at 20$^\circ$
zenith angle, for energies between 3 TeV and 30 TeV.
Histogram: new algorithm, points: conventional analysis.}
\end{center}
\end{figure}

The new analysis technique was applied to the data collected with the
HEGRA telescope system during the extraordinary 1997 flares
of Mkn 501. 
The telescope hardware, the event samples,
the event reconstruction and event selection,
and the determination of energy spectra are described in
Aharonian et al. (1999b), and in Paper I.
In particular, relatively loose cuts on the shower direction and 
in the image shapes are used to select gamma rays from Mkn 501,
in order to minimize the efficiency corrections and the systematic
errors resulting from these corrections, and events are only
used up to core distances of 200~m from the central telescope.
In the current analysis, the new procedure is employed 
to determine the shower
energy on the basis of the measured height of the shower maximum,
and the core location is determined given the known source
location. 
We will, in the following, only address the energy range
above 3 TeV, where the effective detection area of the telescope system
is approaching a constant and where threshold effects are negligible.
The look-up tables used in the energy determination were optimized
for this energy range.
At lower energies, errors in the determination of the energy spectra
are dominated by the modeling of trigger efficiencies (see Paper I),
and no attempt was made to refine the spectra in this range. 
According to Monte Carlo simulations, the improved energy estimate
provides an energy resolution of 10\% to 12\% and has a systematic
deviation between the true energy and the average reconstructed
energy of less than
$\pm$3\%, in the range between  3~TeV and 30~TeV. Fig.~\ref{fig_eres}
compares the performance of this energy reconstruction with the previous
technique; the new approach provides both better resolution and 
reduced non-Gaussian tails.

\section{Tests of the reconstruction algorithm}

A key issue is of course whether the Monte-Carlo simulation 
(Konopelko et al. 1999b) -- from which
the energy look-up tables are derived -- is accurate enough to guarantee
that the identical energy resolution is obtained for the simulations and
the real data. Lacking a monoenergetic test beam, the energy resolution
of Cherenkov telescopes cannot be determined directly. Fortunately,
however, the information provided by the up to four views of a given
shower is highly redundant, and allows a number of cross checks, some
of which are briefly summarized in the following. We address in sequence the
reconstruction of the shower core, the determination of the shower height,
and the energy determination and influence of the atmosphere.

In most of the events in the range above 3 TeV, all four telescopes
trigger and provide images; the mean number of telescopes used in
the reconstruction
is 3.7. The precision in the location of the shower core can be 
tested by reconstructing, in four-telescope events,
the same shower independently by two stereo
subsystems of two telescopes each. The rms difference in the $x$ or $y$
coordinates of the core, comparing two telescopes with the other two,
is about 3.1~m for small core distances, and grows to 5.6~m for
distances of 100~m. These differences are about 30\% larger than
predicted by the simulation, indicating remaining alignment errors.
The same simulation predicts an error in the core coordinates
of less than 2~m rms over the full distance range, when combining
all telescopes. Scaled up by 30\%, this error
is, however, still safely below values which could affect the energy
determination.

\begin{figure}[htb]
\begin{center}
\mbox{
\epsfxsize7.5cm
\epsffile{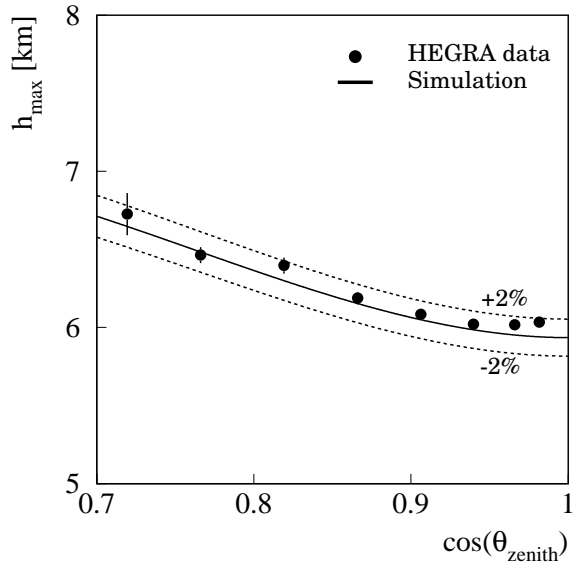}}
\label{fig_hmax}
\caption{Points: average reconstructed shower height for gamma-ray
showers as a function
of zenith angle, for events with at least two images above
100 photoelectrons. Full line: Monte Carlo simulations. Dashed lines:
$\pm$2 \% range around simulation results. }
\end{center}
\end{figure}

For the energy determination, it is crucial that the simulation
provides a proper modeling of the atmosphere, and in particular
that the average height of the shower maximum is reproduced. Otherwise,
the energy correction based on the height of the shower maximum 
might introduce a serious bias. Fig.~\ref{fig_hmax} shows, as a 
function of zenith angle, the measured shower height, compared with
simulations. In order not to introduce a bias by cuts on the
-- potentially biased -- shower energy, events were selected by
requiring two telescopes with at least 100 photoelectrons.
Data and simulation agree to better than 2\%. Simulations show
that even a 5\% mismatch in the average shower height has an almost
negligible effect on the energy resolution, and introduces a bias
in the overall energy scale of less than 3-4\%. Also the widths
of the distributions of shower heights agree within
statistical errors between data and simulation;
at $20^\circ$ zenith angle, e.g., the measured (Gaussian) width is 
$0.82 \pm 0.02$ km, compared to $0.80 \pm 0.01$ km in the simulation.  
We note that due to the selection condition used for the data
of Fig.~\ref{fig_hmax}, the energy of gamma rays
increases with increasing zenith angle. The zenith-angle dependence
of the average shower height is caused by a combination of two effects:
at larger angles (and fixed energy), the shower maximum moves up since
the amount of radiation lengths traversed at a given height increases
like $1/\cos(\theta_{\mbox{zenith}})$; however, showers at larger zenith angles
require a higher energy to provide two images above 100 photoelectrons
and therefore penetrate deeper.

Checks of the energy resolution use the feature that for a given
shower and a core location determined on the basis of the
orientation of the images, the light yield observed in each individual
telescope can be converted into an energy estimate. For the
final energy value these estimates are averaged, but the consistency
of the up to four energy values for each shower provides stringent
tests of the quality of the reconstruction procedure.

A crucial input in the energy reconstruction is the dependence
of  the measured light yield on the distance to the shower core. 
This input can be checked
by comparing the energy estimates of two telescopes located at
different distances from the shower core. We find that the
(averaged) ratio of the energy estimates provided by two telescopes
shows a systematic variation with
the two core distances (in the range between 0 and 200~m), at
a level of 4.2\% rms. A 4.2\% systematic variation should not
spoil the 10\% energy resolution; furthermore, the effect is fully
reproduced by the simulations, which show, under the same 
conditions, a 4.4\% rms variation.

The previous test checks if the simulations reproduce the variation
of the mean light yield with core distance. An equally important
test is to see if the fluctuations are properly described.
Using four-telescope events, the median rms spread of the four individual
energy estimates was determined to $12.3 \pm 0.3\%$, slightly
below the Monte Carlo value of 13.1\%. 
The conventional energy reconstruction shows a $19.5 \pm 0.5\%$
median rms spread between telescopes, compared to the simulation value
of 19.2\%; shower height fluctuations result in strong correlations
between the individual telescopes, and govern the energy resolution.

Finally, one has to worry about varying atmospheric conditions,
which might introduce time-dependent variations in the light yield.
Since such effects will most likely influence all telescopes in
more or less the identical manner, they would not show up in
inter-telescope comparisons. One can distinguish two effects:
long-term seasonal variations and short-term variations.
The influence of seasonal variations on the Cherenkov light yield
is discussed by Bernl\"ohr (2000). Atmospheric density profiles
and temperature profiles on the Canary Islands are continuously monitored
by radiosondes; on the basis of these data (Cuevas 1997) and
of the simulations of Bernl\"ohr (2000) we
conclude that the seasonal variation of the Cherenkov light yield
should be less than 2\% rms for the Mkn 501 data set. More critical
are short-term variations in atmospheric transmission. As discussed
in Paper I, the data set used here is selected to exclude nights
with excessive attenuation. To monitor the attenuation, the trigger
rate of the telescope system is used; the rate varies roughly like
the 1.6th power of the energy threshold and variations in the transmission
cause corresponding rate variations, amplified by a factor 1.6.
For the data set after selection, the mean trigger rate (at a given zenith
angle) varies day-by-day by 4\% rms for the first period of data,
and by 6\% rms for periods 2 and 3 (Aharonian et al. 1999b;
by mistake, the paper quotes 4\% FWHM instead of the correct value
of 4\% rms). These rate fluctuations translate into fluctuations
in the light yield -- and hence of the energy scale -- of 3\% to 4\% rms,
sufficiently small compared to the resolution of 10\% to 12\% rms.

On the basis of these and other tests, we believe that in the data
there are no systematic effects or additional fluctuations, which
would significantly worsen the energy resolution compared to the
resolution predicted by the simulations.

Concerning the absolute energy scale, the new analysis is exactly
like the previous analyses susceptible to imperfections in the
modeling of the atmosphere, of the transmission of the telescope
optical systems, and of the photomultiplier and electronics
gains. Therefore, the 15\% uncertainty on the energy scale
discussed in Paper I also
applies here. The same is true for systematic errors at high
energies due to photomultiplier nonlinearities or electronics saturation.

\section{Results and interpretation}

\begin{figure}
\begin{center}
\mbox{
\epsfxsize8.5cm
\epsffile{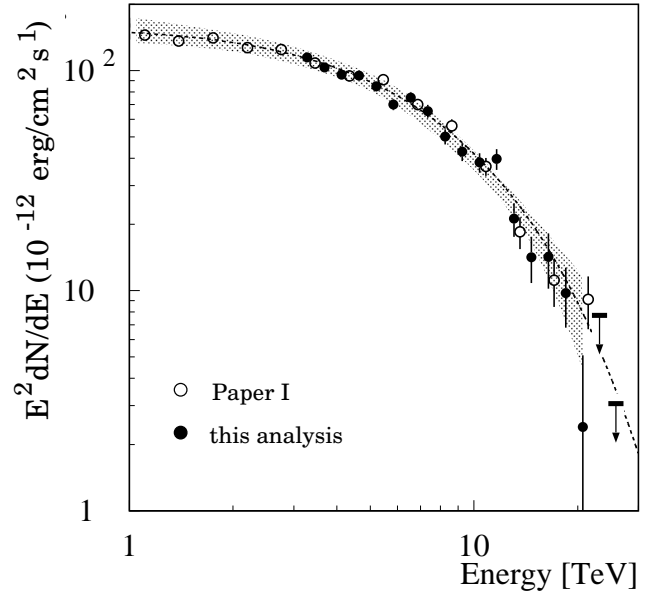}}
\label{fig_spect}
\caption{Spectral energy distribution $E^2$d$N$/d$E$. 
Full symbols and upper limits:
new analysis with improved energy resolution; open symbols: 
conventional analysis, see Paper I. The shaded regions
indicate
systematic errors. In addition, there is a 15\% uncertainty on
the absolute energy scale. The dashed line  shows the fit
$E^{-1.92}\exp(-E/E_o$) with $E_o = 6.2$~TeV to the data of Paper I.}
\end{center}
\end{figure}

Fig.~\ref{fig_spect} shows the spectral energy distribution
obtained with the new energy reconstruction, together with
the previous results of Paper I. The new data points are
listed in Tab.~\ref{tab_dat}.
\begin{table}
\begin{center}
\begin{tabular}{|r|r|r|}
\hline
$E$ [TeV] & $E^2$d$N$/d$E$ & stat. error \\
\hline
  3.28  &    114.7   &   4.6    \\
  3.68  &    103.4   &   4.3    \\
  4.13  &    95.8    &  4.3    \\
  4.63  &    94.8    &  4.4    \\
  5.20  &    84.8    &  4.3    \\
  5.83  &    70.1    &  4.1    \\
  6.54  &    75.4    &  4.5    \\
  7.34  &    65.4    &  4.3    \\
  8.24  &    50.0    &  4.0    \\
  9.25  &    42.8    &  3.9    \\
  10.38  &    38.2   &   3.9    \\
  11.64  &    39.6   &   4.2    \\
  13.06  &    21.2   &  3.7    \\
  14.66  &    14.1   &   3.3    \\
  16.45  &    14.2   &  4.0    \\
  18.45  &    9.7    &  2.9    \\
  20.71  &    2.4    &  2.6  \\
\hline  
\end{tabular}
\caption{Spectral energy distribution $E^2$d$N$/d$E$ in units of
$10^{-12}$ erg~cm$^{-2}$s$^{-1}$. See text for systematic
errors.}
\label{tab_dat}
\end{center}
\end{table}
Except for the highest point in the previous data set,
at $E$ = 21.45~TeV, the two analyses are in very good agreement.
In Paper I, the influence of the energy resolution was
addressed by studying a model, where the source spectrum shows a power
law with an index around 2, an exponential cutoff, and in addition
a sharp cutoff at an energy $E_{\mbox{cut}}$. At the $2\sigma$ level,
a hard cutoff below 16 TeV was excluded in Paper I; this result implies
that at the $2\sigma$ level, in particular the data point at 21.45~TeV
is consistent with spill-over from lower energies. In the new
analysis, no significant signal is observed in this energy range,
and upper limits are below the value of the previous point.
We conclude therefore that at these very highest energies, the
spectrum appears indeed steeper than indicated by the data of 
Paper I, but consistent within the systematics quoted in Paper I.
While this change goes in the expected direction, a statistical
fluctuation at the 2-3$\sigma$ level cannot be excluded as the 
origin of the change, due to the low
statistics for this data range in either analysis.
In the range up to the data point at 17.08 TeV of Paper I, there
is almost no change observed due to the improved energy resolution.
Combining the previous data below 3 TeV with the new data
points above 3 TeV, a good fit to the data is achieved
with the parametrization
$$
\mbox{d}N/\mbox{d}E \sim E^{-2.03 \pm 0.03} 
\exp\left(-(E/E_o)^{1.31 \pm 0.11}\right)
$$
with $E_o = 8.21 \pm 0.63$ TeV.
The fit indicates that indeed a super-exponential cutoff is
favored. We note -- as already mentioned in Paper I -- that the
fit parameters are strongly correlated and cannot be varied
independently within their errors. Errors quoted for the fit
parameters are purely statistical.

Carrying out a similar analysis with a hard cutoff for the new
results, a cutoff below 16.5 TeV can be excluded.

In summary, the conclusions concerning the level of the infrared
background allowed by the HEGRA Mkn 501 energy spectra remain
unchanged: taken at face value, the HEGRA data and the infrared
flux levels of Finkbeiner et al. (2000) are inconsistent; 
making full use of the
systematic errors allowed for the HEGRA data and in particular for the infrared
flux determinations, the two data sets can be brought to a marginal
consistency, without a clear need to invoke exotic explanations.

\section*{Acknowledgments}

The support of the HEGRA experiment by the German Ministry for Research 
and Technology BMBF and by the Spanish Research Council
CYCIT is acknowledged. We are grateful to the Instituto
de Astrof\'{i}sica de Canarias for the use of the site and
for providing excellent working conditions. We gratefully
acknowledge the technical support staff of Heidelberg,
Kiel, Munich, and Yerevan. GPR acknowledges
receipt of a Humboldt Foundation postdoctoral fellowship.

\end{document}